# ON ORBIT DIMENSIONS UNDER A SIMULTANEOUS LIE GROUP ACTION ON $n$ COPIES OF A MANIFOLD.

MIREILLE BOUTIN

ABSTRACT. We show that the maximal orbit dimension of a simultaneous Lie group action on $n$ copies of a manifold does not pseudo-stabilize when $n$ increases. We also show that if a Lie group action is (locally) effective on subsets of a manifold, then the induced Cartesian action is locally free on an open subset of a sufficiently big (but finite) number of copies of the manifold. The latter is the analogue for the Cartesian action to Ovsiannikov's theorem on jet spaces and is an important fact relative to the moving frame method and the computation of joint invariants. Some interesting corollaries are presented.

## 1. INTRODUCTION

The moving frame method [9, 10] provides an algorithmic way to compute invariants of Lie group actions on manifolds. A necessary and sufficient condition for the existence of a (local) moving frame is that the action be (locally) free. If this is not the case, one can consider the induced action on the $k^{th}$ order jet space and try to compute a moving frame on this bigger space. A theorem by Ovsiannikov [11] (as corrected by Olver [8]) states that if a group acts (locally) effectively on subsets, then there exists an integer $k_0$ such that the prolonged action of the given group is locally free on an open and dense subset of the $k_0^{th}$ order jet space. Such a prolongation leads to the computation of differential invariants.

Another way to prolong the action is to let the group act on many copies of the manifold (Cartesian action). In many cases, it has been observed that considering the action on two, three, four, ... copies of the manifold eventually leads to an action that is locally free on an open subset of the manifold. The main goal of this paper is to guarantee that for group actions which are (locally) effective on subsets this will always be the case, so a local moving frame always exists. Such a prolongation leads to the computation of joint invariants.

Besides direct implications concerning joint invariants [5], joint differential invariants [3, 6] and numerically invariant numerical algorithms [7] , this result is important in computer vision, in particular for curve recognition modulo Lie group action [1, 2].

## 2. About Stabilization of the Cartesian action

Let $G$ be a Lie group acting on a $m$-dimensional manifold $M$. Let $M^{\times(k)} := M \times M \times \ldots \times M$ ($k$ times) be the Cartesian product of $k$ copies of $M$. The action of $G$ on $M$ induces an action of $G$ on $M^{\times(k)}$, namely $g \cdot (x_1, \ldots, x_k) = (g \cdot x_1, \ldots, g \cdot x_k)$ for $g \in G$ and $x_1, \ldots, x_k \in M$.

**Definition 2.1.** We say that a Lie group action on a manifold $M$ is *semi-regular* if all the orbits have the same dimension. If in addition we have that for every point $x \in M$ there exists an arbitrarily small neighborhood $U$ such that the intersection of $U$ with the orbit through $x$ is connected, then we say that the action is *regular*.

**Definition 2.2.** We say that a real valued function $I : U \subset M \to \mathbb{R}$ is an *invariant* if $I(g \cdot x) = I(x)$, for all $g \in G$ and all $x \in U$. We say that a real valued function $I : U \subset M \to \mathbb{R}$ is a *local invariant* if there exists a neighborhood $N$ of $e \in G$ such that $I(g \cdot x) = I(x)$, for all $g \in N$ and all $x \in U$.

The results we will derive are based on the following important theorem. See [4] for a proof.

**Theorem 2.3.** *If $G$ acts on an open set $O \subset M$ semi-regularly with $s$ dimensional orbits, then $\forall x_0 \in O$ there exist $m - s$ functionally independent local invariants $I_1, \ldots, I_{m-s}$ defined on a neighborhood $U$ of $x_0$ such that any other local invariant $I$ defined near $x_0$ is a function $I = f(I_1, \ldots, I_{m-s})$. If the action of $G$ is regular, then the local invariants can be taken to be invariants in a neighborhood of $x_0$, and two points $x_1, x_2 \in U$ are in the same orbit if and only if $I_i(x_1) = I_i(x_2)$, $\forall i = 1, \ldots, m - s$.*

The set of invariants $I_1, \ldots, I_{m-s}$ is often called a *complete fundamental set of invariants*.

Denote by $s_k$ the maximal orbit dimension of the Cartesian action of $G$ on $M^{\times(k)}$, and by $\mathfrak{M}^k$ the set of points $z \in M^{\times(k)}$ for which the orbit through $z$ has maximal dimension $s_k$. Observe that the action of $G$ on $\mathfrak{M}^k$ is semi-regular, for any $k$. Observe also that $\mathfrak{M}^k$ is open, for any $k$.

**Definition 2.4.** The minimal integer $n_0$ such that $s_n = s_{n_0}$, for all $n \geq n_0$ is called the *stabilization order*. We call $s_{n_0}$ the *stabilization dimension*.

**Example 2.5.** Let the special Euclidean group $SE(2)$ act on the plane by rotating and translating the points in the standard way. More precisely, for $g \in SE(2)$ and $x \in \mathbb{R}^2$, let

$$\bar{x} = g \cdot x = Rx + b, \text{ with } R = \left( \begin{array}{cc} \cos\theta & \sin\theta \\ -\sin\theta & \cos\theta \end{array} \right) \text{ and } b \in \mathbb{R}^2.$$

The group $SE(2)$ has dimension three while the plane is a two dimensional manifold. In fact, the Euclidean action on one copy of the plane is transitive so the dimension of the orbit through any point is equal to two.

If we consider the Cartesian action of $SE(2)$ on two copies of the plane, namely

$$g \cdot (x_1, x_2) = (Rx_1, Rx_2) + (b, b)$$

for $R$ and $b$ as defined above, then the dimension of the space acted on is four while the dimension of the orbits is two for the points on the diagonal $D_2 := \{(z, z) | z \in \mathbb{R}\}$ and three everywhere else. Since the maximal orbit dimension $s_2$ is equal to the dimension of the group, then for all $n \geq 2$, $s_n$ is also equal to the dimension of the group. Therefore two is the stabilization order and three is the stabilization dimension.

**Example 2.6.** Consider the projective group $PSL(3)$ acting on the plane as

$$\bar{x} = g \cdot x = \frac{Ax + b}{c^\dagger x + d}, \text{ with } \left( \begin{array}{cc} A & b \\ c^\dagger & d \end{array} \right) \in GL(3)$$

for $g \in GL(3)$ and $x \in \mathbb{R}^2$. We have $\dim PSL(3) = 9$ while $\dim \mathbb{R}^2 = 2$ is equal to the dimension of the orbits. In fact, this action is transitive. Also for $n = 2, 3, 4$, the Cartesian action of $PSL(3)$ is transitive on $\{(x_1, \ldots, x_n) \in M^{(n)} | x_1, \ldots, x_n \text{ are distinct }\}$. However, the dimension of the orbits is bounded by eight because $PSL(3)$ contains a one-dimensional subgroup that acts trivially on the plane. Therefore the stabilization order is four and the stabilization dimension is eight.

**Definition 2.7.** If there exists an integer $n$ such that $s_n = s_{n+1} < s_{n+2}$, then we say that the Cartesian group action *pseudo-stabilizes* at order $n$.

The next lemma states that Cartesian group actions do not pseudo-stabilize. Note that the analogue for the action on the jet space is not true.

**Lemma 2.8.** *If $s_{n-1} < s_n = s_{n+1}$ then $n$ is the stabilization order.*

*Proof.* Let $p \in \mathfrak{M}^n$. By Theorem (2.3), there exists $\{I_1, \ldots, I_{nm-s_n}\}$ a complete fundamental set of invariants in a neighborhood $U_n \cap \mathfrak{M}^n$ of $p$. Of course, $nm - s_n$ might be zero but that is not relevant to our proof. Let $\bar{I}_i(z_1, \ldots, z_{n+1}) = I_i(z_1, \ldots, z_n)$, for $i = 1, \ldots, nm - s_n$. Since $s_n = s_{n+1}$ and the invariants $\bar{I}_1, \ldots, \bar{I}_{nm-s_n}$ are functionally independent on an open set $U_{n+1} \cap \mathfrak{M}^{n+1}$, there exist $m$ invariants $J_1, \ldots, J_m : U_{n+1} \subset M^{\times(n+1)} \to \mathbb{R}$ such that

$$\{\bar{I}_1(z_1, \ldots, z_{n+1}), \ldots, \bar{I}_{nm-s_n}(z_1, \ldots, z_{n+1}), J_1(z_1, \ldots, z_{n+1}), \ldots, J_m(z_1, \ldots, z_{n+1})\}$$

is a complete fundamental set of invariants on an open set $\tilde{U}_{n+1} \cap \mathfrak{M}^{n+1}$. Define $\tilde{J}_i : \mathfrak{M}^{n+2} \to \mathbb{R}$ by

$$\tilde{J}_i(z_1, \ldots, z_{n+2}) = J_i(z_1, \ldots, z_n, z_{n+2}),$$

for $i = 1, \ldots, m$. Since $\bar{I}_1, \ldots, \bar{I}_{nm-s_n}, J_1, \ldots, J_m$ are functionally independent and $\bar{I}_1, \ldots, \bar{I}_{nm-s_n}$ only depend on the first $n$ points $z_1, \ldots, z_n \in M$, then the invariants

$$\begin{aligned}
\bar{I}_1(z_1, \ldots, z_{n+1}), &\quad \ldots \quad, \bar{I}_{nm-s_n}(z_1, \ldots, z_{n+1}) \\
J_1(z_1, \ldots, z_{n+1}), &\quad \ldots \quad, J_m(z_1, \ldots, z_{n+1}) \\
\tilde{J}_1(z_1, \ldots, z_{n+2}), &\quad \ldots \quad, \tilde{J}_m(z_1, \ldots, z_{n+2})
\end{aligned}$$

are functionally independent on an open set $U_{n+2} \cap \mathfrak{M}^{n+2}$.

By Theorem (2.3) the number of functionally independent $(n+2)$-point joint invariants in a neighborhood of $p$ is equal to $(n+2)m - s_{n+2}$. Therefore

$$\begin{aligned}
(n+2)m - s_{n+2} &\geq nm - s_n + 2m \\
\Leftrightarrow s_{n+2} &\leq s_n.
\end{aligned}$$

But $s_{n+2} \geq s_n$, so we conclude that $s_{n+2} = s_n$. We can repeat the same argument many times to show that $s_k = s_n$ for all $k \geq n$, and therefore $n$ is the order of stabilization. $\qquad \square$

Let $r$ be the dimension of $G$.

**Corollary 2.9.** *If $n_0$ is the stabilization order of the Cartesian action, then $n_0 \leq r - s_1 + 1 \leq r + 1$.*

*Proof.* By Lemma (2.8), we have

$$\begin{aligned}
s_2 &\geq s_1 + 1 \\
s_3 &\geq s_1 + 2 \\
&\;\;\vdots \\
s_{n_0} &\geq s_1 + n_0 - 1.
\end{aligned}$$

(1)

But $s_{n_0} \leq r$, therefore $n_0 \leq r - s_1 + 1$. In addition, since $s_1 \geq 0$, we can also say that $n_0 \leq r + 1$. $\qquad \square$

## 3. About local freeness and effectiveness of the Cartesian action

Let $S$ be a subset of $M$.

**Definition 3.1.** The *isotropy subgroup* of $S$ is the set

$$G_S = \{g \in G \text{ such that } g \cdot S = S\}$$

The *global isotropy subgroup* of $S$ is the set

$$G_S^\star = \{g \in G \text{ such that } g \cdot z = z, \text{ for all } z \in S\}$$

**Definition 3.2.** We say that $G$ acts on $M$ *effectively* if $G_M^\star = \{e\}$. We say that $G$ acts on $M$ *locally effectively* if $G_M^\star$ is a discrete subgroup of $G$.

The action of the special Euclidean group described in Example 2.5 is an example of effective action, while the action of the projective group described in Example 2.6 is not effective.

**Definition 3.3.** We say that $G$ acts on $M$ *effectively on subsets* if, for any open subset $U \subset M$, $G_U^\star = \{e\}$. We say that $G$ acts on $M$ *locally effectively on subsets* if, for any open subset $U \subset M$, $G_U^\star$ is a discrete subgroup of $G$.

Observe that (local) effectiveness on subsets implies (local) effectiveness. However, the converse is not true as illustrated by the following example (kindly provided by professor P.J. Olver.)

**Example 3.4.** Let $h : \mathbb{R} \to \mathbb{R}$ be the function defined by

$$h(x) = \left\{ \begin{array}{ll} 0 & \text{if } x \leq 0 \\ e^{-\frac{1}{x}} & \text{if } x > 0 \end{array} \right. .$$

Consider the action of the plane on itself given by

$$(\bar{x}, \bar{y}) = (x, y + uh(x) + vh(-x)),$$

for $u, v \in \mathbb{R}$. This action is effective since the only trivial transformation is $(u, v) = (0, 0)$. However, it is not effective on subsets since it is not effective on $\{x \in \mathbb{R} | x > 0\}$ for example.

The previous example corresponds to a smooth action that is *not* analytic. For analytic actions, (local) effectiveness implies (local) effectiveness on subsets.

**Lemma 3.5.** *A Lie group $G$ acts on a subset $S \subset M$ (locally) effectively if and only if the Cartesian action of $G$ on $S^{\times(n)}$ is (locally) effective.*

*Proof.* Since

$$g \cdot z = z, \text{ for all } z \in S$$
$$\Leftrightarrow g \cdot (z_1, z_2) = (z_1, z_2), \text{ for all } z_1, z_2 \in S,$$

then $G_S^\star = G_{S^{\times(n)}}^\star$. Therefore

$$G_S^\star = \{e\} \Leftrightarrow G_{S^{\times(n)}}^\star = \{e\}$$

and also

$$G_S^\star \text{ is a discrete subgroup } \Leftrightarrow G_{S^{\times(n)}}^\star \text{ is a discrete subgroup } .$$

$\square$

Recall that $r$ is the dimension of the Lie group $G$. We now give a necessary and sufficient condition for the maximal orbit dimension of a prolonged Lie group action to reach $r$.

**Theorem 3.6.** *A Lie group $G$ acts (locally) effectively on subsets of $M$ if and only if for any open subset $S \subset M$, the stabilization dimension of the action of $G$ on $S$ is equal to the dimension of $G$.*

*Proof.* If $G$ does not act (locally) effectively on subsets of $M$, then by Lemma (3.5) $G$ does not act (locally) effectively on subsets of $M^{\times(n)}$, for any $n \in \mathbb{N}$. Therefore the orbit dimension through any point in any open subset $S_n \subset M^{\times(n)}$ is strictly smaller than $\dim G$ for all $n \in \mathbb{N}$, which proves the necessity of the (local) effectiveness on subsets.

To prove sufficiency, suppose that $G$ acts (locally) effectively on subsets of $M$ and suppose there exists an open subset $S \subset M^{\times(n)}$ such that the stabilization order $n$ of the action of $G$ on $S$ is $s_n < r$. Let $\mathfrak{S}^n$ be the set of points $z^{(n)} \in S$ such that the orbit through $z^{(n)}$ has maximal dimension $s_n$. The set $\mathfrak{S}^n$ is open, so for all $z^{(n)} \in \mathfrak{S}^n$ there exists a neighborhood $N(z^{(n)}) \subset \mathfrak{S}^n$.

Let $v_1, \ldots, v_r$ be a basis for the Lie algebra of $G$ and let $v_1^{(i)}, \ldots, v_r^{(i)}$ be the corresponding vector fields on $M^{\times(i)}$. Assume that $v_1^{(n)}, \ldots, v_{s_n}^{(n)}$ are linearly independent on $N(z^{(n)})$.

If $s_n < r$, we can write

$$v_{s+k}^{(n)}\Big|_{(z_1, \ldots, z_n)} = \sum_{i=1}^{s_n} A_k^i \, v_i^{(n)}\Big|_{(z_1, \ldots, z_n)}, \text{ for all } k = 1, \ldots, r - s_n,$$

where $A_k^i = A_k^i(z_1, \ldots, z_n)$. Since the same relationships hold on $M^{\times(n+1)}$, we can also write

$$v_{s+k}^{(n+1)}\Big|_{(z_1, \ldots, z_{n+1})} = \sum_{i=1}^{s_n} A_k^i \, v_i^{(n)}\Big|_{(z_1, \ldots, z_{n+1})}, \text{ for all } k = 1, \ldots, r - s_n,$$

with the same coefficients $A_k^i$.

We have

$$v_j^{(n+1)}\Big|_{(z_1,\dots,z_{n+1})} = \left(v_j^{(1)}\Big|_{(z_1)}, \dots, v_j^{(1)}\Big|_{(z_{n+1})}\right) \text{ for } j = 1, \dots, r.$$

In particular, for $k = 1, \dots, r - s$

$$v_{s+k}^{(n+1)}\Big|_{(z_1,\dots,z_{n+1})} = \left(v_{s+k}^{(1)}\Big|_{(z_1)}, \dots, v_{s+k}^{(1)}\Big|_{(z_{n+1})}\right),$$

$$\text{but also } v_{s+k}^{(n+1)}\Big|_{(z_1,\dots,z_{n+1})} = \sum_{i=1}^{s_n} A_k^i \left(v_i^{(1)}\Big|_{(z_1)}, \dots, v_i^{(1)}\Big|_{(z_{n+1})}\right).$$

Therefore

$$v_{s+k}^{(1)}\Big|_{(z_j)} = \sum_{i=1}^{s_n} A_k^i\, v_i^{(1)}\Big|_{(z_j)}, \text{ for } j = 1, \dots, n+1.$$

If we fix $z_1 = z_1^0, \dots, z_n = z_n^0$ in $M$, then all $A_k^i$'s are constants, which contradicts the effectiveness of the action of $G$ on open subsets of $M$. Therefore $s_n = r$. □

Examples 2.5 and 2.6 illustrate this result.

**Definition 3.7.** We say that $G$ acts *freely* on $M$ if for all $z \in M$, $G_z = \{e\}$. If for all $z \in M$ the set $G_z$ is a discrete subgroup of $G$, then we say that $G$ acts *locally freely* on $M$.

**Corollary 3.8.** *If a Lie group $G$ acts (locally) effectively on subsets of $M$, then for all $n$ greater than or equal to the stabilization order $n_0$, $G$ acts on the open set $\mathfrak{M}^n$ locally freely and $\mathfrak{M}^n$ is dense.*

One may ask whether it is possible to replace local freeness by freeness in the previous statement. The answer is unfortunately negative, as illustrated by the following counterexample (kindly provided by professor P.J. Olver.)

**Example 3.9.** Consider the action of the real line on the plane given in polar coordinates by

$$(\bar{r}, \bar{\theta}) = (r, \theta + rt),$$

for $t \in \mathbb{R}$. This action is locally free and effective. But for all $n \in \mathbb{N}$, the Cartesian action is not free on the dense subset

$$\left\{(r_1, \theta_1, r_2, \theta_2, \dots, r_n, \theta_n) \in (\mathbb{R}^2)^{\times(n)} \Big| \frac{r_i}{r_1} \in \mathbb{Q} \text{ for } i = 2, \dots, n\right\}.$$

## 4. Some interesting corollaries

Let $v_1, \ldots, v_r$ be a basis for the Lie algebra of $G$ and let $v_1^{(i)}, \ldots, v_r^{(i)}$ be the corresponding vector fields on $M^{\times(i)}$. Let $z = (x_1, \ldots, x_m)$ be local coordinates for $M$ and write $v_k^{(1)} = \sum_{i=1}^{m} \xi_k^i(x) \frac{\partial}{\partial x^i} = \sum_{i=1}^{m} \xi_k^i(x) \partial_i$ for $k = 1, \ldots, r$. We define the analogue of the Lie matrix of order $n$ for the jet space as

$$\tilde{L}_n(z_1, \ldots, z_n) = \begin{pmatrix} \xi_1^1(z_1), & \cdots & , \xi_1^m(z_1), & \xi_1^1(z_2), & \cdots & , \xi_1^m(z_2), & \cdots & , \xi_1^m(z_n) \\ \vdots & & \vdots & \vdots & & \vdots & & \vdots \\ \xi_r^1(z_1), & \cdots & , \xi_r^m(z_1), & \xi_r^1(z_2), & \cdots & , \xi_r^m(z_2), & \cdots & , \xi_r^m(z_n) \end{pmatrix}$$

Since the number of functionally independent infinitesimal generators of the group action at a certain point is equal to the dimension of the orbit through this point, we have the following lemma.

**Lemma 4.1.** *The orbit through $(z_1, \ldots, z_n) \in M^{\times(n)}$ has dimension equal to the rank of the matrix $\tilde{L}_n(z_1, \ldots, z_n)$.*

Therefore if a group action stabilizes at order $n_0$, then

$$s_{n_0} = \max_{z_1, \ldots, z_{n_0} \in M} \{ \text{ rank } \tilde{L}_{n_0}(z_1, \ldots, z_{n_0}) \}.$$

**Corollary 4.2.** *The action of $G$ on $M$ is locally effective on subsets if and only if for all open subsets $S \subset M$, there exists $n \in \mathbb{N}$ such that the rank of the matrix $\tilde{L}_n(z_1, \ldots, z_n)$ is equal to $r$ for some $z_1, \ldots, z_n \in S$.*

Combining with corollary (2.9) we get the following.

**Corollary 4.3.** *The action of $G$ on $M$ is locally effective on subsets if and only if for all open subsets $S \subset M$, the rank of the matrix $\tilde{L}_n(z_1, \ldots, z_{r+1})$ is equal to $r$ for some $z_1, \ldots, z_{r+1} \in S$*

Combining with lemma (2.8), we have the following result.

**Lemma 4.4.** *A group action stabilizes at order $n_0$ if and only if*

$$\max_{z_1, \ldots, z_{n_0} \in M} \{ \text{ rank } \tilde{L}_{n_0}(z_1, \ldots, z_{n_0}) \} = \max_{z_1, \ldots, z_{n_0+1} \in M} \{ \text{ rank } \tilde{L}_{n_0+1}(z_1, \ldots, z_{n_0+1}) \}$$

For $n \in \mathbb{N}$, define the projection $\pi(\mathfrak{M}^n) = \{ z_1 \in M \mid \exists z_2, \ldots z_n \in M \text{ with } (z_1, \ldots, z_n) \in \mathfrak{M}^n \}$.

In the case of an action prolonged on the jet space, one studies the notion of singular points (see [8]), which are nothing but points whose jet of arbitrarily large order does not belong to an orbit of maximal dimension. The following facts tells us that the analogue of singular points for Cartesian group actions does not exist.

**Lemma 4.5.** *Let $n_0$ be the order of stabilization of a group action. Then $M = \pi(\mathfrak{M}^{n_0+1})$.*

*Proof.* If $z_1 \in \pi(\mathfrak{M}^{n_0})$, then $z_1 \in \pi(\mathfrak{M}^{n_0+1})$. Now if $z_1 \notin \pi(\mathfrak{M}^{n_0})$, then consider $(z_2, \ldots, z_{n_0+1}) \in \mathfrak{M}^{n_0}$. We have $(z_1, \ldots, z_{n_0+1}) \in \mathfrak{M}^{n_0+1}$ so $z_1 \in \pi(\mathfrak{M}^{n_0+1})$. $\qquad\square$

Considering $\det \tilde{L}$, we obtain the analogue to an important result [8] concerning the Lie determinant.

**Proposition 4.6.** *If $\tilde{L}(z_1, \ldots, z_n)$ is a square matrix, then the equation*

$$\det \tilde{L}(z_1, \ldots, z_n) = 0$$

*is invariant under $G$.*

*Proof.* A point $(z_1, \ldots, z_n) \notin \mathfrak{M}^n$ if and only if the determinant of $\tilde{L}_n(z_1 \ldots, z_n)$ is zero. But if $(z_1, \ldots, z_n) \notin \mathfrak{M}^n$ then $g \cdot (z_1 \ldots, z_n)$ is also not in $\mathfrak{M}^n$, for all $g \in G$. Therefore

$$\det \tilde{L}(g \cdot z_1, \ldots, g \cdot z_n) = 0 \text{ for all } g \in G.$$

$\qquad\square$

More generally, we have

**Proposition 4.7.** *The set of points*

$$\{(z_1, \ldots, z_n) | \ rank \ \tilde{L}(z_1, \ldots, z_n) = k\}$$

*is invariant under $G$.*

Let $X$ be a $p$-dimensional manifold.

**Definition 4.8.** Given a set of $r$ vector-valued functions $f_1, \ldots, f_r : X \to \mathbb{R}^q$ we say that they are *linearly dependent on a subset $W \in X$* if there exists a non-trivial relationship of the form $\sum_{k=1}^{r} c_k f_k(x) \equiv 0$, with $c_1, \ldots, c_r \in \mathbb{R}$, which holds for all $x \in X$. In the negative, we say that they are *linearly independent*.

We now define the analogue of the $n^{th}$ order Wronskian matrix at $x_1, \ldots, x_n \in X$

$$\tilde{W}_n(x_1, \ldots, x_n) = \begin{pmatrix} f_1^1(x_1), & \ldots & , f_1^q(x_1), & f_1^1(x_2), & \ldots, & f_1^q(x_2), & \ldots, & f_1^q(x_n) \\ \vdots, & & \vdots & \vdots & & \vdots & \vdots \\ f_r^1(x_1), & \ldots & , f_r^q(x_1), & f_r^1(x_2), & \ldots, & f_r^q(x_2), & \ldots, & f_r^q(x_n) \end{pmatrix}$$

**Theorem 4.9.** *A set of of $r$ vector-valued functions $f_1, \ldots, f_r : X \to \mathbb{R}^q$ is linearly independent on subsets of $X$ if and only if their $(r+1)^{st}$ order Wronskian $W_{r+1}(x_1, \ldots, x_{r+1})$ has maximal rank $r$ in a dense subset of $X^{(r+1)}$.*

*Proof.* Let $M = X \times \mathbb{R}^q$ and write $z \in M$ as $z = (x, v)$, with $x \in X$ and $v \in \mathbb{R}^q$. Consider the following action of $G = \mathbb{R}^r$ on the $m$ dimensional manifold $M$.

$$z \mapsto \left( x, v + \sum_{k=1}^{r} t_k f_k(x) \right), \text{ for } t = (t_1, \ldots, t_r) \in \mathbb{R}^r.$$

The vector fields

$$\mathbf{v}_k = \sum_{l=1}^{m} f_k^l(x) \frac{\partial}{\partial v^l}, \text{ for } k = 1, \ldots, r$$

are the infinitesimal generators of this group action.

The functions $f_1, \ldots, f_r$ are linearly independent on subsets if and only if the vectors $\mathbf{v}_1, \ldots, \mathbf{v}_r$ are linearly independent on subsets. The latter is true if and only if $G$ acts locally effectively on subsets of $M$. By Corollary 4.3, this happens if and only if for any open subset $S \subset M$, the maximal orbit dimension of $G$ action on $S^{\times(r+1)}$ is equal to $r$, which is true if and only if

$$\max_{z_1, \ldots, z_{r+1} \in S} \text{rank } \tilde{L}_{r+1}(z_1, \ldots, z_{r+1}) = r,$$

for any open $S \subset M$. But

$$\tilde{L}_n(z_1, \ldots, z_n) = \begin{pmatrix} 0, & \ldots & ,0, & f_1(x_1), & 0, & \ldots & ,0, & f_1(x_2), & \ldots & ,f_1(x_n) \\ \vdots & & \vdots & \vdots & \vdots & & \vdots & \vdots & & \vdots \\ 0, & \ldots & ,0, & f_r(x_1), & 0, & \ldots & ,0, & f_r(x_2), & \ldots & ,f_r(x_n) \end{pmatrix},$$

and

$$\tilde{W}_n(x_1, \ldots, x_n) = \begin{pmatrix} f_1(x_1), & f_1(x_2), & \ldots & ,f_1(x_n) \\ \vdots & \vdots & & \vdots \\ f_r(x_1), & f_r(x_2), & \ldots & ,f_r(x_n) \end{pmatrix}.$$

Therefore the rank of $\tilde{L}_{r+1}(z_1, \ldots, z_n)$ is equal to the rank of $\tilde{W}_{r+1}(x_1, \ldots, x_n)$. So $f_1, \ldots, f_r$ are linearly independent on subsets if and only if

$$\max_{x_1, \ldots, x_{r+1} \in U} \text{rank } \tilde{W}_{r+1}(x_1, \ldots, x_{r+1}) = r,$$

for any open $S \subset U$, which is equivalent to saying that

$$\text{rank } \tilde{W}_{r+1}(x_1, \ldots, x_{r+1}) = r,$$

on a dense subset of $X$.                                              $\square$

## Acknowledgments

I want to thank my advisor Peter J. Olver for many useful suggestions, for his advice and for his support.